\documentclass[
prl,twocolumn,preprintnumbers,superscriptaddress,showpacs
]{revtex4-1}
\usepackage{graphicx,subfigure}
\usepackage{amsmath,amssymb,amsfonts}
\usepackage{bm}
\usepackage{color}
\usepackage{comment}
\usepackage{slashed} 
\usepackage{lineno}

\newcommand\sect[1]{\section{#1}}

\def \be {\begin{equation} }
\def \ee {\end{equation}}
\def \bea {\begin{eqnarray}}
\def \eea {\end{eqnarray}}
\def \bem {\begin{multline}}
\def \eem {\end{multline}}

\def \bes {\begin{subequations} }
\def \ees {\end{subequations}}

\def \pd {\partial}

\def \a {\alpha}
\def \b {\beta}

\def \e {\epsilon}
\def \g {\gamma}

\def \o {\omega}

\def \l {\lambda}

\def \<{\langle}
\def \>{\rangle}
\def \+{\dagger}
\def \({\left(}
\def \){\right)}

\def \[{\left[}
\def \]{\right]}

\def \vq {\bm{q}}
\def \vx {\bm{x}}

\def \vo {\bm{\o}}

\def \vj {\bm{j}}

\def \vB {\bm{B}}
\def\vo{\bm{\o}}
\def \vE {\bm{E}}

\def \vv {\bm{v}}

\def \hB {\hat{\vB}}
\def \he {\hat{e}}

\def \tr {\tilde{r}}

\def \CME {{\rm CME}}

\newcommand{\sgn}{ {\rm sgn} }
\newcommand{\prj}{ {\mathcal P} }

\newcommand{\gam}{ \gamma }

\newcommand{\LLL}{ {\rm LLL} }
\newcommand{\V}{ {\scriptscriptstyle \rm V} }

\newcommand{\q}{ q_{ \scriptscriptstyle f} }

\newcommand{\ext}{ {\scriptscriptstyle \rm ext} }
\newcommand{\R}{ {\! \scriptscriptstyle  R} }

\def\tr{\mbox{tr}}

\def\sgn {\textrm{sgn}}

\newcommand{\bo}{{\bm{\omega}}}
\newcommand{\bB}{{\bm{B}}}
\newcommand{\bS}{{\bm{S}}}

\newcommand{\Bhat}{\hat {\bm{B}}}

\usepackage[normalem]{ulem}  % \sout{old text} for strikeout

\renewcommand\sout{\bgroup \color{red} \ULdepth=-.5ex \ULset}
\graphicspath{{./plots/}}

\begin{document}

\title{
%Anomalous charge redistribution induced by magneto-vorticity coupling in a chiral fluid
%in the strong magnetic field limit: 
%\\
%Charge redistribution from magneto-vorticity coupling in anomalous hydrodynamics
Charge Redistribution from Anomalous Magnetovorticity Coupling
}

\author{Koichi~Hattori}
\affiliation{Physics Department, Fudan University, Shanghai 200433, China}
\affiliation{RIKEN-BNL Research Center, Brookhaven National Laboratory, Upton,
 New York 11973-5000, U.S.A
}
%\affiliation{Nishina Center, RIKEN, Wako, Saitama 351-0198, Japan
%}

\author{Yi Yin}
\affiliation{Department of Physics,
Brookhaven National Laboratory, Upton, New York 11973-5000, USA}

\date{ \today}

\begin{abstract}
%We illustrate a novel transport phenomena in a chiral fluid.
%We argue the energy spectrum of chiral fermions in the lowest Landau level (LLL) are shifted due to spin-rotation coupling and consequently the presence of both vorticity and the magnetic field would induce local charge redistribution in such chiral fluid.
%We will support our findings by evaluating the corresponding Kubo formula using diagrammatical techniques and argue the magnitude of this effect is tied to quantum anomaly. 
%
%reveals new anomalous transport phenomena. 
%Based on an analysis by the Kubo formula, the corresponding transport coefficients 
%are tied to anomaly and are shown to be independent of temperature or density correction. 
%Our analysis by the  suggests that the corresponding transport coefficients 
%are tied to  and are protected from temperature or density correction. 
We investigate novel transport phenomena 
%\com{``Phenomena'' is not a singular form. Cannot come with article ``a''.} 
in a chiral fluid originated from an interplay between a vorticity and strong magnetic field, 
which induces a redistribution of vector charges in the system and an axial current along the magnetic field. 
The corresponding transport coefficients are obtained from 
an energy-shift argument for the chiral fermions in the lowest Landau level (LLL) due to a spin-vorticity coupling 
and also from diagrammatic computations on the basis of the linear response theory.
Based on consistent results from both methods, 
we observe that the transport coefficients are proportional to the anomaly coefficient 
and are independent of temperature and chemical potential. 
We therefore speculate that these transport phenomena are connected to quantum anomaly. 
%and could be universal.  
%\com{``Phenomena'' is not a singular form.}
%\com{What do you mean ``universal'' more specifically?}
%suggest that 
%the transport coefficients are independent of temperature and density, 
%and are connected . 
%\com{`Phenomena' is a plural form. `Would' is not usually used in abstract. The first and second sentences were redundant. 
%`thus' is not a conjunction.}
%We support our finding by considering an adiabatic energy-shift of the lowest Landau level (LLL) due to spin-rotation coupling and by a a diagrammatic computation using the Kubo formula. 
%We also argue that the corresponding transport coefficients are tied to chiral anomaly thus are independent of temperature and density
%Based on an adiabatic level-shift argument in the presence of a spin-rotation coupling 
%and a diagrammatic computation by the Kubo formula, 
%we argue that corresponding transport coefficients are independent of temperature and density. 
\end{abstract}

\maketitle
\preprint{RBRC-1167}

%\linenumbers
\sect{Introduction} 
A number of intensive and extensive studies have shown that 
%systems with chiral fermion exhibit anomalous transport phenomena induced by the quantum anomaly. 
the dynamics of chiral fermions in various systems manifests itself in anomalous transport phenomena
induced by the quantum anomaly. 
The broad set of such systems includes the primordial electroweak plasma 
in the early Universe \cite{Giovannini:1997eg}, the QCD matter created in the relativistic heavy-ion collisions 
\cite{Kharzeev:2007jp,*Kharzeev:2007tn,*Liao:2014ava}, 
and newly invented condensed matter systems - Weyl and Dirac semimetals \cite{ciudad2015weyl, semimetal} 
(see also Refs.~\cite{Kharzeev:2015znc, MS, rev-Huang} for recent reviews). 

%A chiral medium (parity-violating) in the presence of magnetic field and/or vorticity 
%exhibits many interesting transport phenomena associated with quantum anomaly. 
%The broad set of systems with those anomalous transport phenomena is widely discussed in the literature and includes the primordial electroweak plasma in early universe (see e.g. \cite{Giovannini:1997eg}), the QCD matter created in heavy-ion collisions 
%(see e.g. \cite{Kharzeev:2007jp,*Kharzeev:2007tn,*Liao:2014ava}) and newly discovered condensed matter systems - Weyl and Dirac semi-metals~(see e.g. \cite{ciudad2015weyl, semimetal}). 

One prominent example of such anomalous transport phenomena is known as 
the Chiral Magnetic Effect (CME) \cite{Vilenkin:1980fu,Kharzeev:2012ph,*Kharzeev:2013ffa}, 
that is, an induction of a vector (electric) current in response to a magnetic field $\vB$. 
In the presence of a chirality imbalance quantified by the axial chemical potential $\mu_{A}$, 
the vector current is induced along $\vB$ as 
\begin{equation}
\label{CME}
\vj_{V,\CME}= q_f  C_{A}\mu_{A}\vB\, ,
\end{equation}
where $ q_f$ is the electric charge of the chiral fermion
and $C_{A}= 1/2\pi^2$ % \com{[$ N_c $ removed.]}
is the nonrenormalizable coefficient characterizing the chiral anomaly relation 
\begin{equation}
\label{anomaly}
\pd_{\mu}j^{\mu}_{A} = q_f^2 C_{A} \vE\cdot \vB \, .
\end{equation}
The CME current has been investigated by various theories and methods
%from weakly coupled fermion gas~\cite{} to gauge theories at infinite `t Hooft coupling~\cite{}, 
that consistently confirm Eq.~(\ref{CME}) (see Refs.~\cite{Kharzeev:2013ffa,Kharzeev:2015znc} for reviews). 
This indicates the universality of CME attributed to the topological nature of the chiral anomaly. 

It is also known that the magnetic field induces not only the vector current but also an axial current. 
Namely, the Chiral Separation Effect (CSE)~\cite{Son:2004tq} emerges in the presence of a vector chemical potential $\mu_{V}$ as 
\begin{equation}
\label{CCSE}
\vj_{A, {\rm CSE}} = q_f  C_{A} \mu_{V}\vB\, .
\end{equation}

A vorticity in a chiral fluid plays a similar role as that of the magnetic field, 
and hence induces anomalous vector and axial currents---this is 
referred to as the Chiral Vortical Effect (CVE) \cite{PhysRevD.20.1807, Kharzeev:2007tn,Banerjee:2008th,Erdmenger:2008rm}. 
The CME and CVE have been understood on equal footing 
within the framework of anomalous hydrodynamics from the second law of thermodynamics \cite{Son:2009tf}. 

%\changed{
It should be emphasized that the above studies are devoted to the separate effects 
of the magnetic field $ \bB$ or the vorticity $ \bo$. 
In the pioneering hydrodynamic analysis with the anomaly~\cite{Son:2009tf}, 
both vorticity and magnetic field are accounted as the first order in the gradient expansion. 
Consequently, the coupling between $ \bB$ and $ \bo$ is dropped 
as a higher-order effect in that systematic framework. 
However, in the context of magnetohydrodynamics, 
the magnetic field is not screened in a medium, 
and its strength can be much larger than the gradients, 
suggesting the importance of going beyond the conventional gradient expansion. 
%}
%\changed{
%The conventional studies are dedicated to the separate effects 
%of the magnetic field $ \bB$ or the vorticity $ \bo$ on the basis of the derivative expansion. 
%However, a magnetic field is not screened in a medium 
%and can be much larger than the scale of derivatives. 
%Therefore, %we need to go beyond the conventional counting scheme, and 
%%the coupling term between the vorticity and magnetic field 
%a strong magnetic field enhances their coupling term $ \bB\cdot \bo$, 
%which becomes comparable to the first-order terms in the derivative expansion. 
%%In this letter, we investigate this strong-field regime, and show that 
%%new anomalous transport phenomena are induced by their interplay. 
%}

In this Letter, we will show that the interplay between the vorticity and strong magnetic field 
induces a local vector-charge density 
\begin{equation}
\label{jV}
\Delta j^{0}_{V}=  q_f \frac{C_{A}}{2} \( \vB\cdot\vo\)\, , 
\end{equation}
%\com{1/2 included.}
where the vorticity is defined by $\vo=\frac{1}{2}\nabla\times \vv$. 
Below, Eq.~(\ref{jV}) will be consistently derived both from 
an analysis of the energy shift by a spin-vorticity coupling in the lowest Landau level (LLL) 
and from a diagrammatic computation on the basis of the Kubo formula. 
Remarkably, 
$\Delta j^{0}_{V}$ in Eq.~\eqref{jV}
%the coefficient in Eq.~\eqref{jV} 
is proportional to anomaly coefficient $C_{A}$, and does not depend on temperature and chemical potential. 
This suggests a connection to the underlying quantum anomaly as discussed below.

It is worth pointing out that Eq.~\eqref{jV} does not create a globe vector charge, i.e., $\int d^{3}\vx\, \Delta j^{0}_{V}=0$.
This can be seen as 
$\int \! d^{3}x \vB\cdot \vo 
%= \int \! d^{3}x \, \vB\cdot (\nabla\times \vv ) 
= \frac{1}{2}\,\int \! d^{3}x \, \nabla \cdot (\vv \times  \vB) 
= \frac{1}{2}\,\int_{\scriptscriptstyle \partial V} \! d{\bm S} \cdot  (\vv\times\vB) = 0$ 
for a homogenous magnetic field $\vB$ .
As usual, we assume that the flow velocity $\vv$ vanishes sufficiently fast at the asymptotic region. 
Therefore, Eq.~\eqref{jV} indicates a redistribution of the vector charge in the system. 
In general, due to the inherent inhomogeneity of the vorticity, 
Eq.~\eqref{jV} will induce intriguing charge distribution patterns in a chiral fluid.  %as illustrated in Fig.~\ref{}. 
%in accord with the total charge conservation. %\com{ Shall we add a figure?}
%
%and does not cause a generation of the total charge. 
%Indeed, integrating over 
%
%and the boundary condition for an open system that 

%\com{Is this change ok?}
%In other particular boundary conditions, 
%\com{Or we have used the boundary condition for magneto-hydro}.
%
%Due to the inherent inhomogeneity of the vorticity, we note that the vector charge accumulates only locally 
%and the currents (\ref{results}) are given in the differential forms. 
%To find global charge distribution, one needs to invoke the hydrodynamic framework with specific boundary conditions 
%which might provide intriguing distribution patterns in bulk and at the boundary. 
%For a simple boundary condition with a vanishing fluid velocity $\vv \to 0$ near the boundary, 
%we find that the net charge is conserved as 
%
%
%\begin{figure}[b]
%\centering
%%\includegraphics[width=0.4\textwidth]{./plot/LLL.pdf}
%\caption
%{
%\label{fig:rotation}
%\com{[Maybe, one more figure.]}
%Fluid vortices in a strong magnetic field.
%}
%\end{figure}

We will also show that, accompanying the induction of the local vector charge imbalance (\ref{jV}), 
a new contribution to the axial current emerges as% \com{[Position of the hat changed.]}
\begin{equation}
\label{jA}
\Delta\vj_{A}=  \vert q_f \vert \frac{C_{A}}{2} \( \vB\cdot\vo\) \hB\, ,
\end{equation}
%\com{1/2 included.}
where $ \hat \vB = \vB/\vert\vB\vert$ is the unit vector along the magnetic field. 
This is an analogue of CSE \eqref{CCSE} induced by the imbalance of vector charge $\mu_{V}$. 
Here, it is remarkable that the axial current is dynamically generated without an initial finite value of $\mu_{V}$. 

The generation of the vector charge density in chiral media is also discussed in condensed matter physics 
on the basis of the realization of an effective axial gauge field \cite{2013PhRvB..87w5306L,Ebihara:2015aca}. 
However, to the best of our knowledge, Eqs.~\eqref{jV} and \eqref{jA} are new in the literature. 
%\sout{We point out, for the first time to the best of our knowledge, 
%that the vorticity plays the role of the effective axial gauge field. }
Since the vorticity is one of the most important dynamical variables in magnetohydrodynamics, 
its coupling to the strong magnetic field, indicated by Eqs.~\eqref{jV} and \eqref{jA}, 
should be incorporated in anomalous magnetohydrodynamics (see also Eq.~\eqref{hydro}). 
Results reported in this Letter clearly open a new avenue for studying the intriguing interplay 
occurring in a wide variety of chiral media in strong magnetic fields. 
%in any chiral medium that can be described by chiral magneto-hydrodynamics.  

%\begin{figure}
%\centering
%\includegraphics[width=0.4\textwidth]{./plot/LLL.pdf}
%\caption
%{
%\label{fig:LLL}
%LLL
%}
%\end{figure}

%\begin{figure}[t]
%\centering
%\includegraphics[width=0.45\textwidth]{./plot/sea.pdf}
%\caption
%{
%\label{fig:LLL}
%Adiabatic energy shifts by the spin-rotation coupling 
%across the surface of bottomless sea. 
%Spins of right and left handed particles in the LLL 
%are aligned in the same direction.
%}
%\end{figure}

%\section{Physical picture}

\sect{Physical picture} 

Prior to performing an explicit diagrammatic analysis, 
we first provide a physical picture as to why the vorticity would 
induce a local vector-charge density when coupled to a magnetic field.  

We shall consider chiral fermions in the presence of a static and homogeneous magnetic field $\vB$. 
The energy spectra of chiral fermions are discretized into the Landau levels (LLs). 
We next turn on a slowly varying velocity field $\vv$ which leads to a nonzero vorticity $\vo = \frac{1}{2}\nabla\times \vv$.
%As long as $\o \ll \sqrt{|q_{f} B|}$ with $q_{f}$ being the electric charge of fermions, 
%we only need to consider effects which are linear in $\o$. \com{Why?}
After a sufficiently long time, each fluid cell reaches a local equilibrium with the single-particle distribution function 
given by $f(\epsilon, \omega) =f_{0}(\e')$ where $f_{0}$ denotes the equilibrium distribution function. 
Our key observation is that the vorticity shifts the single-particle energy from $\e$ to $\e'$ by an amount 
$ \Delta \e \equiv \e^\prime - \e = - \bS \cdot \bo$. 
Here, $\bS$ is the intrinsic angular momentum (spin) carried by fermions.
Such an energy shift due to the spin-vorticity coupling can be derived 
by observing the shift of the single-particle Hamiltonian in a rotating frame~\cite{Mashhoon:1988zz, *Hehl:1990nf}. 
The energy shift also naturally arises in the equilibrium fermion distribution %of particles with spin 
by computing the distribution function which maximizes the entropy~\cite{PhysRevD.20.1807, Becattini:2013fla} 
or by working out a constraint imposed by the detailed balance~\cite{Chen:2014cla}. 
%Since the higher LLs are degenerated with respect to the spin states, 
%the effects of the spin-vorticity coupling on the higher LLs can be neglected 
In the every higher LL, the spin-vorticity coupling splits the degenerated spin states into the opposite directions, 
%In the higher LLs, since the spin-vorticity coupling shifts the energy levels of 
%the pairwise spin states in the opposite directions, 
so that these effects cancel at the linear order in $\vo$ when averaging over the spin. 
We will therefore concentrate on the unique grand state, i.e., the lowest Landau level (LLL). 

In the LLL, the spin directions of both right- and left-handed particles 
are frozen in the same direction along the magnetic field $\bS_{R/L} = \frac{1}{2}  \sgn(q_f) \Bhat$, 
and those of antiparticles are oriented in the opposite direction. 
Consequently, the energy shift in the LLL has no dependence on the chirality and is given by 
\begin{equation}
\label{e-omega}
\Delta \epsilon_{LLL}^{\pm} =\mp \frac{1}{2} \sgn(q_{f})\, \Bhat \cdot \bo\, , 
\end{equation}
%where the signs ``$-$'' and ``$+$'' 
where the upper and lower signs refer to a particle and antiparticle, respectively. 
Below, we take $\vB=B\, \hat{e}_{3}$ without loss of generality. 

We are now ready to compute the change of the density of chiral fermions $n_{R/L}$ due to the vorticity. 
As explained above, we only need to consider the contributions from the LLLs 
where the fermion dynamics is reduced to the (1+1) dimensional one along $ \bB$. 
Expanding $f_{0}(\e')$ up to the linear order in $\Delta \e$, and 
using the linear dispersion relation of the right-handed LLL fermion, i.e., $\e_{\rm LLL}= + p^{3}$, 
%with $+$ ($-$) corresponding to the right (left)-handed fermions, 
we find 
%\com{Signs etc were corrected below. Please check again.}
\begin{eqnarray}
\label{Delta-nR}
\Delta n_{R}&=&
% \(\frac{|q_{f}\, B|}{2\pi}\)\, \changed{ \(-\frac{1}{2} \sgn(q_{f})\, \Bhat \cdot \bo \) }
% \nonumber \\
% &\times&
%\Big[\, \int^{\infty}_{0}\, \frac{dp^{3}}{2\pi}\, 
%\frac{\pd\, f_{0}(p^{3})}{\pd p^{3}}
%-
%\int^{0}_{-\infty}\, \frac{dp^{3}}{2\pi}\, 
%\frac{\pd\, \bar{f}_{0}(p^{3})}{\pd p^{3}} 
%\, \Big ]\, 
 \(\frac{|q_{f}\, B|}{2\pi}\) 
\Big[\, \Delta \e_{LLL}^{+} \int^{\infty}_{0}\, \frac{dp^{3}}{2\pi}\, 
\frac{\pd\, f_{0}(p^{3})}{\pd p^{3}}
 \nonumber \\
 && \hspace{2cm}
 +
 \Delta \e_{LLL}^{-}
\int^{0}_{-\infty}\, \frac{dp^{3}}{2\pi}\, 
\frac{\pd\, \bar{f}_{0}(p^{3})}{\pd p^{3}}
\, \Big ]\, 
\nonumber \\
&=& q_f \frac{C_{A} }{4} ( \bB \cdot \bo ) \, \[\,  f_{0}(0)+\bar{f}_{0}(0)\, \]
\nonumber\\
&=& q_f \frac{C_{A} }{4} (\bB \cdot \bo )
\, .
\end{eqnarray}
Here, the factor of $|q_{f}B|/2\pi$ is the density of states in the LLL per unit transverse area. 
The Fermi-Dirac distribution functions of particles and antiparticles are given by 
$f_{0}(\e)= 1/[e^{(\e-\mu)/T}+1] $ and $ \bar{f}_{0}(\e)= 1/[e^{-(\e-\mu)/T}+1]$, respectively. 
We have used the fact that $f_{0}(\infty)=\bar{f}_{0}(-\infty)=0$.
Remarkably, one finds an identity 
 $f_{0}(0)+\bar{f}_{0}(0)=1$, which is independent of temperature $T$ and chemical potential $\mu$.
Consequently, the last line in Eq.~\eqref{Delta-nR} is also independent of $T$ and $\mu$. 
For the left-handed fermions with $\e_{\rm LLL}= - p^{3}$, 
a similar computation leads to $\Delta  n_{L}=\Delta n_{R}$.  
Therefore, we find $\Delta n_{V}=\Delta  n_{L} + \Delta n_{R} = q_f C_{A}  (\bB \cdot \bo )/2$. 
This is the aforementioned result shown in Eq.~\eqref{jV}.

%Furthermore, since the right (left)-handed chiral fermions in the LLL 
%are moving along $\he_{3}$ with the speed of light $1$ ($-1$), 
Furthermore, since the chiral fermions in the LLL are moving along $\he_{3}$ with the speed of light, 
the generation of $\Delta n_{R, L}$ also induces currents 
$\Delta j^{3}_{R}= \sgn(S_R) \Delta n_{R} $ and $ \Delta j^{3}_{L}= - \sgn(S_L)  \Delta n_{L}$. 
%Note that the directions of the chirality flow are determined by the combination 
%of the spin direction $\sgn(S_{R/L})$ and the sign of $ \Delta n_{R/L}$. 
Therefore, from Eq.~\eqref{Delta-nR}, we find an axial current 
$\Delta j^{3}_{A}=\Delta j^{3}_{R}-\Delta j^{3}_{L} = \vert q_f \vert C_{A}  (\bB \cdot \bo ) \Bhat /2$. 
This verifies Eq.~\eqref{jA}. 
On the other hand, the vector current vanishes $\Delta j^{3}_{V}=\Delta j^{3}_{R} + \Delta j^{3}_{L} = 0 $. 
Alternatively, 
one might also interpret the amount of the energy shift~\eqref{e-omega} as an effective chemical potential 
$ \Delta  \mu_{R,L} =  - \Delta \e_{LLL}^+ $ %which is also understood as the generation of charge density 
(see also Ref.~\cite{Chen:2015hfc} for a discussion on the analogy between rotating and charge density). 
Plugging the effective vector chemical potential $ \Delta \mu_V = (\Delta  \mu_R +\Delta   \mu_L)/2 = \sgn(q_f) \bo \cdot \Bhat$ 
into the CSE current \eqref{CCSE}, 
we again find the generation of the axial current~\eqref{jA} along the magnetic field. 
Note that the sign of the axial current depends only on the direction of the vorticity 
and is independent of that of the magnetic field. 

Importantly, 
since Eq.~\eqref{Delta-nR} and thus Eq.~\eqref{jV} 
manifestly depend on the anomaly coefficient $C_{A}$, but neither $T $ nor $ \mu$, 
it is natural to speculate that the form of Eq.~\eqref{jV} is nonrenormalizable and is tied to the chiral anomaly. 
In the subsequent section, we will verify Eqs.~\eqref{jV} and \eqref{jA} by an explicit field-theoretical computation, 
and provide further evidence on the connection to the quantum anomaly.

\begin{figure}[t]
\centering
\includegraphics[width=0.3\textwidth]{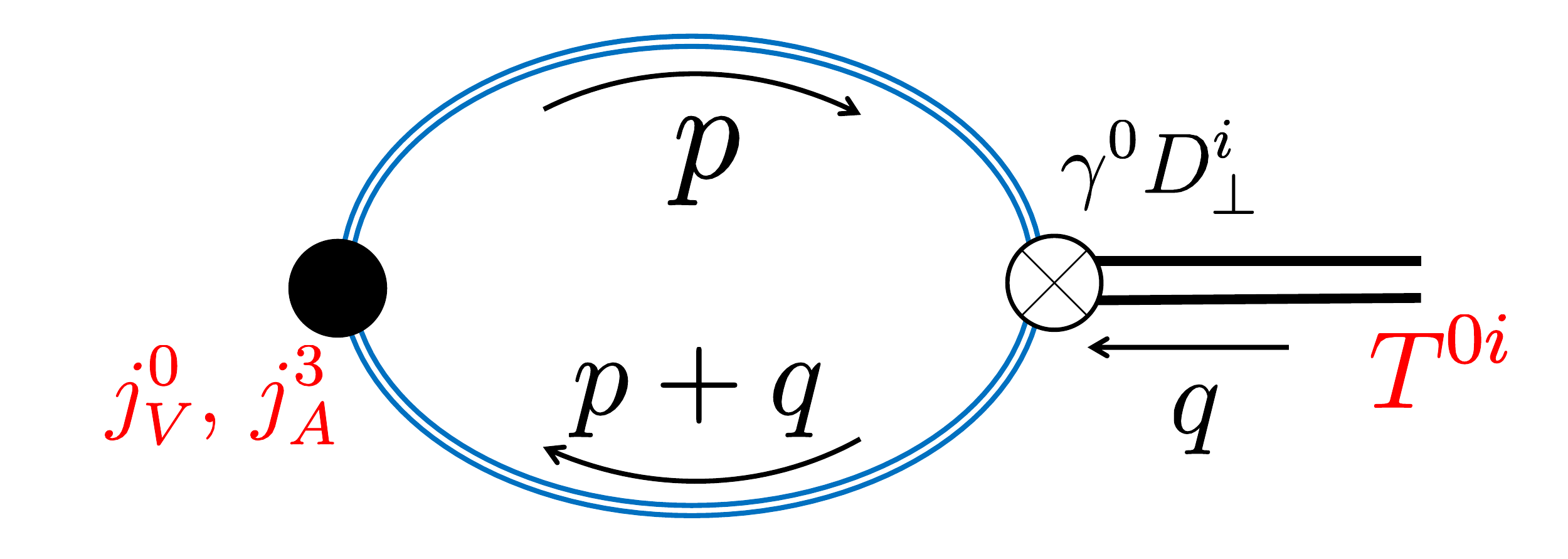}
\caption
{
\label{fig:kubo}
One-loop diagram for the Kubo formula. The internal double lines represent 
the fermions in the LLL.
}
\end{figure}

\sect{Diagrammatical computations} 
We now perform the field-theoretical computation. % by using the linear response theory. 
We will consider the response of the chiral medium to the vorticity $\vo$ in the presence of external magnetic field. 
An inhomogeneous velocity field $\vv(\vx)$ may be mimicked by 
turning on a fictitious gravitational field $ds^{2}=dt^{2}+2 \vv\cdot d\vx dt -d\vx^{2}$, i.e., 
$g_{0i}(\vx)=\delta_{ij} v^{j}(\vx)$. 
Therefore, the Fourier representation of Eq.~\eqref{jV} is cast into 
\begin{equation}
\label{pertub}
j^{0}_{V} = \frac{\l}{2}\, 
\epsilon^{ljk}\, \hB_{l}\, (i q_{j})\, g_{0k}\, , 
%
%\frac{i q_{1}}{2}\, v_{2}\,  , 
\end{equation}
%\com{[Need to think how to include B in lambda.]}
where we used $\vo =\frac{1}{2}\nabla\times \vv$, and 
$\l $ is the transport coefficient to be computed below.  
%\com{{Check the factor.}} 
%Without loss of generality, 
We again take the direction of the magnetic field to be $\vB = B \he_{3}$ 
and specify an inhomogeneous velocity profile as $\vv= v (x_{1}) \he_{2}$ 
or equivalently an inhomogeneous perturbation of the metric as $\delta g_{02}(x_{1})$. 
Inverting Eq.~(\ref{pertub}), we find the Kubo formula 
\begin{equation}
\label{kubo}
\l =(-2i) \lim_{\vq\to 0}\, 
\Bigg[\lim_{\omega \to 0}\, \frac{\pd}{\pd q^{1}}G^{0, 0 2}_{\R} (\o, \vq)
\Bigg]\, , 
\end{equation}
with the retarded Green's function (see Fig.~\ref{fig:kubo}): 
\begin{eqnarray}
\label{GR}
G^{0, 02}_{\R} (x-x^\prime) &\equiv& 
\Big\langle j^{0}_{V} (x) T^{0 2}(x^\prime) \Big\rangle\, \theta(t-t^\prime)\, . 
\end{eqnarray}
A similar Kubo formula was used to study the CVE 
without an external magnetic field in Ref.~\cite{Amado:2011zx,*Landsteiner:2011cp}. 

%In the present case, the effects of the external magnetic field will be included in the fermion propagators. 
%We use the Kubo formula which was constructed to investigate the chiral vortical effect in Ref.~\cite{Landsteiner:2011cp}. 
%While the authors of Ref.~\cite{Landsteiner:2011cp} has considered only the transport properties without an external magnetic field, 
%the general framework can be applied to the present case with an external constant magnetic field: 
%the effects of the external constant magnetic field will be included in the fermion propagators. 
%Following from the derivation presented in Ref.~\cite{}, the relevant Kubo formula is found to be 
%\begin{equation}
%\label{Kubo}
%\xi^{\mu i}_{\chi} =\frac{1}{2 i} \lim_{\vq\to 0}\lim_{\omega \to 0} \sum_{\ell,m}
%\Bigg[\frac{\epsilon^{i \ell m}}{q^{m}} G^{\mu, 0 \ell}_{\R, \chi} (\o, \vq)
%\Bigg]
%\, ,
%\end{equation}
%with 
%%\com{[A factor of i was removed.]}
%where $\chi=V, A$ labels the channel and $ \epsilon^{i \ell m}$ is the antisymmetric tensor $( \epsilon^{123} = +1)$. 

We now evaluate the Green's function \eqref{GR} in a weak coupling theory.
The vector current and energy-momentum tensor of Dirac fermions are given by 
%\com{Axial current is included.}
\bes
\begin{eqnarray}
\label{current}
%&&
j^{\mu}_\V (x) &\equiv& \bar \Psi (x) \gam^\mu \Psi (x)
%, \ \ 
%j_\A (x) = \bar \Psi (x) \gam^\mu \gam^5  \Psi (x)
,
\\
\label{T0i}
%&&
T^{0 i}(x) &\equiv&
\frac{i}{2} \bar \Psi(x) ( \gam^0 D^i + \gam^i D^0)  \Psi(x)\, , 
\end{eqnarray}
\ees
where $g^{\mu\nu} = {\rm diag} (1,-1,-1,-1) $ and $\gam^5 \equiv i \gam^0\gam^1\gam^2\gam^3$. 
The covariant derivative $D^\mu = \partial^\mu + i \q A^\mu_\ext(x)$ 
includes the gauge potential $A^\mu_\ext(x)$ for the magnetic field $ \vB$. 
%To be specific, we will use Landau's gauge $\vA_{\ext} = B\, x_{1}\, \he_{2}$ 
%below.%but it is clear that the final results should be gauge independent. 

We consider the one-loop diagram %sketched in Fig.~\ref{fig:kubo} 
composed of the dressed Fermion propagators in the external magnetic field. 
%\com{----------}
Below, we will restrict ourselves to the contributions from the lowest Landau levels (LLLs).
This is because the anomalous currents such as the CME current are solely 
transported by the fermions populated in the LLL. 
%We observe that for anomalous transport in the presence of magnetic field. 
%the contribution is saturated by those from . 
%For example, 
%it is well-known that CME is due to LLL: the contributions from HLL cancel out.
%For this reason,
%we will further simplify our computation by considering the contributions from LLL only. 
%\com{----------Merge this paragraph with the following.}
%
%In the strong field limit, the dynamics of the fermions populated 
%in the LLLs is essentially (1+1)-dimensional. 
%%and their spins have the definite direction along the magnetic field. 
%Therefore, only the longitudinal components of the currents are relevant 
%and the other transverse components are vanishing. 
%Moreover, when the magnetic field and the vorticity are perpendicular to each other, 
%\com{[Comment on this configuration.]}
%This can be seen by decomposing the quark field as $\Psi = \psi_\LLL \prj_+$ 
%with the LLL wave function $\psi_\LLL$ and 
%the spin-projection operator $\prj_+=(1 + i s_f \gam^1 \gam^2 ) /2$ with $s_f = \sgn(\q B)$, 
%assuming that the magnetic field is oriented in the third direction. 
%One finds $\prj_+ \gam^\mu \prj_+ = \prj_+ \gam^\mu_\parallel$ 
%where the longitudinal gamma matrix is defined by $\gam_\parallel^\mu = (\gam^0,0, 0,\gam^3)$. 
%Also, because of an identity $\gam^5 \gam_\parallel^\mu \prj_- 
%= - s_f \epsilon_\parallel^{\mu\nu} \gam_{\parallel \nu} \prj_-$, 
%with the antisymmetiric tensor $\epsilon_\parallel^{03} = - \epsilon_\parallel^{03} = +1 $, 
We therefore project the fermion wave function into the LLL: $ \Psi= \prj_+\,\psi_\LLL $ 
with $\psi_\LLL$ and $\prj_{\pm}=(1 \pm i s_f \gam^1 \gam^2 ) /2$ being 
the LLL wave function and the spin-projection operator with $s_f \equiv \sgn(\q B)$, respectively. 
%\begin{eqnarray}
%T^{0i} 
%%&=& \frac{i}{2} \bar \psi_\LLL \prj_- ( \gam^0 D^i + \gam^i D^0) \prj_- \psi_\LLL
%%\\
%&=& \frac{i}{2} \bar \psi_\LLL  \gam^0 D^i_\perp \prj_- \psi_\LLL 
%%\ \ ({\rm only \, when} \, i = 1,2)
%\end{eqnarray}

The coordinate representation of the retarded Green's function~\eqref{GR} 
is written as (cf. Fig.~\ref{fig:kubo}):
\begin{eqnarray}
\label{eq:G}
G_{\R}^{0, 0 2} (x-x^\prime) 
%&=& i \langle [j^3_\A (x^\prime) , T^{0i}(x)] \rangle \theta(t^\prime-t)
%\nonumber
%\\
%&=& - \frac{1}{2} \langle \, \bar \psi_\LLL (x^\prime) \gam^5 \gam^3_\parallel \prj_- \psi _\LLL(x^\prime)
%\cdot \bar \psi_\LLL(x)  \gam^0 D^i _\perp \prj_- \psi_\LLL(x) \, \rangle
%\nonumber
%\\
%&=& - \frac{1}{2} (-1) \tr[ \, \gam^5 \gam^3 _\parallel \prj_-
%S_\LLL (x^\prime,x) \gam^0 \left( D^i_\perp  S_\LLL (x,x^\prime)  \right) \, ]
%\nonumber
%\\
&=& \frac{1}{2i}  \tr[ \,  \gam^0  \prj_+S_\LLL (x,x^\prime) 
\nonumber
\\
&& \hspace{1cm} \times 
\gam^0 \left( D^{2}_{\! x^\prime} \,  S_\LLL (x^\prime, x)  \right) \, ]\, ,
%\nonumber
%\\
\end{eqnarray}
%where we denote the transverse component of the covariant derivative as $D_\perp^\mu$. 
where we have used the fact that the second term of Eq.~\eqref{T0i} 
vanishes for the transverse components $(i=1,2) $, when the wave function is projected to the LLL. 
Here, 
$ S_\LLL (x^\prime, x) =\<\psi_{\rm LLL}(x') \bar \psi_{\rm LLL}(x)\>$ 
%(\com{is this definition correct? YY}) 
symbolically represents the LLL propagator in the medium and is factorized as~\cite{Sch, MS} 
%Its explicitly form is not important at this moment. 
%in the vacuum and in finite temperature and/or density. 
%However, we will soon find that only the vacuum part contributes to the transport coefficients 
%and the temperature and/or density correction exactly vanishes. 
%It has been known that the symmetry-breaking part of the propagator 
%in a constant external field is factorized as  
\begin{eqnarray}
S_\LLL (x^\prime,x) = e^{i \phi(x^\prime , x )}\, \tilde{S}_\LLL (x^\prime- x) \, ,
\end{eqnarray}
where the Schwinger phase is given by 
\begin{eqnarray}
\hspace{-0.5cm}
\phi(x^\prime, x) = - q_f \! \int_x^{x^\prime} \!\!\!\!   dz^\mu \!
\left[ A_\mu^ \ext (z) + \frac{1}{2}  F_{\mu \nu}^\ext (z^\nu - x^\nu ) \right]
\, ,
\end{eqnarray}
%where $F_{\mu \nu}^\ext  $ is the field strength tensor for $ A_\mu^ \ext $. 
with  $F_{\mu \nu}^\ext  = \partial_\mu A_\nu^ \ext  - \partial_\nu  A_\mu^ \ext $. 
The above integrand is curl free, and, hence, the integral is path independent. 
%A straightforward calculation gives:
%\begin{eqnarray}
%%D^{2}_{x'}\, \[e^{i\phi(x',x)}\, \tilde{S}_{\rm LLL}(x',x)\]
%D^{2}_{x'}\, S_{\rm LLL}(x',x)
%&=&
%e^{i\phi(x',x)}\,\[-\pd_{x'_{2}} + i s_{f}\frac{|q_{f}B|}{2}\(x'_{1}-x_{1}\)\]\, 
% \nonumber\\
%&\times&\tilde{S}_{\rm LLL}(x'-x)\, . 
%\end{eqnarray}
%Therefore, a straightforward calculation gives $ D^\ell_{\! x^\prime} \, \phi (x^\prime,x) =
%\phi(x^\prime , x ) \{ \partial^\ell _{\! x^\prime} - i s_f  \epsilon_\perp^{\ell k} \Delta x_k \vert q_f B \vert/ 2 \}  $,  
%\begin{eqnarray}
%D^i_\perp S_\LLL (x^\prime,x) 
%%&=& D^i_\perp e^{i \phi(x, x^\prime )} S_\LLL (x-x^\prime) 
%%\nonumber
%%\\
%&=&  e^{i \phi(x, x^\prime )} (\partial^i _\perp
%- i \s  \frac{\epsilon_\perp^{ik} \Delta x_k}{2 \lf^2}  ) S_\LLL (x-x^\prime) 
%\nonumber
%\\
%\end{eqnarray}
Therefore, a straightforward calculation gives $ D^\mu_{\! x^\prime} \, \phi (x^\prime,x) =
\phi(x^\prime , x ) \{ \partial^\mu _{\! x^\prime} - i q_f F_\ext^{\mu\nu} \Delta x_\nu /2 \}  $,  
where $\Delta x^\mu =  x^{\prime \mu} - x^\mu $. 
%\com{[4D Covariant form is the simplest.]}
%and $\epsilon_\perp^{12} = - \epsilon_\perp^{21} = +1  $. 
Consequently, the Schwinger phases in Eq.~\eqref{eq:G} cancel each other 
as $\phi(x,x^\prime) + \phi(x^\prime,x)  =0 $. 
The remaining parts then depend only on the difference $ \Delta x^\mu$ 
and are independent of the gauge potential, 
indicating the manifest translational and gauge invariances. 
%
%Thus, observe that $ D^\ell_{\! x^\prime} \, \phi (x^\prime,x) =
%\phi(x^\prime , x ) \{ \partial^\ell _{\! x^\prime} - i s_f  \epsilon_\perp^{\ell k} \Delta x_k/(2 \lf^2) \}  $,  
%%\begin{eqnarray}
%%D^i_\perp S_\LLL (x^\prime,x) 
%%%&=& D^i_\perp e^{i \phi(x, x^\prime )} S_\LLL (x-x^\prime) 
%%%\nonumber
%%%\\
%%&=&  e^{i \phi(x, x^\prime )} (\partial^i _\perp
%%- i \s  \frac{\epsilon_\perp^{ik} \Delta x_k}{2 \lf^2}  ) S_\LLL (x-x^\prime) 
%%\nonumber
%%\\
%%\end{eqnarray}
%where $\Delta x^\mu =  x^{\prime \mu} - x^\mu $ and 
%$\epsilon_\perp^{12} = - \epsilon_\perp^{21} = +1  $. 
%Therefore, after the operation of the covariant derivative, 
%
%
%
%While the gauge and translational invariances are, at this stage, not manifest 
%due to the gauge potential for the external magnetic field, 
%one can show that the Greens's function (\ref{eq:G}) 
%is invariant with respect to these symmetries in the following way. 
%

With these manifest symmetries, we are now ready to transform \eqref{eq:G} into the Fourier space:
%\begin{eqnarray}
%\label{G-Fourier}
%G^{0,02}_{R}(q_{\parallel};q_{\perp}) &=&
%\int \frac{d^{2}p_{\perp}}{(2\pi)^{2}}\, 
%\int\frac{d^{2}p_{\parallel}}{(2\pi)^{2}}\, 
%\tr [\, \g^{0} \prj_{+}\, \tilde{S}_{\rm LLL}(p_{\parallel}+q_{\parallel}; p_{\perp}+q_{\perp})
%\nonumber\\
%&\times& 
%\(-p_2+ i s_{f}\frac{|q_{f}B|}{2}\pd_{p_{1}}\)\, \g^{0}\, \tilde{S}_{\rm LLL}(p_{\parallel}; p_{\perp})
%]\, . 
%\end{eqnarray}
\begin{eqnarray}
\label{G-Fourier}
\hspace{-0.5cm}
G^{0,02}_{R}(q) &=&
\int \frac{d^{4}p}{(2\pi)^{4}}\, 
%\int\frac{d^{2}p_{\parallel}}{(2\pi)^{2}}\, 
\tr \big[\, \g^{0} \prj_{+}\, \tilde{S}_{\rm LLL}(p+q)
\nonumber
\\
&& \hspace{0.5cm}
\times
\big(p^2 + i s_{f}\frac{|q_{f}B|}{2} \frac{\partial \, }{\partial p^1} \big) \g^{0}\, \tilde{S}_{\rm LLL}(p)
\, \big] \, . 
\end{eqnarray}
%\com{[$ p^0$ is not $\omega $.]}
%\com{[Maybe, subscripts are ok as the notation is written just below.]}
Note that $\tilde{S}_{LLL}$ is completely factorized 
into the transverse and longitudinal parts as \cite{MS, FHYY}: 
\begin{eqnarray}
\tilde{S}_\LLL (p_\parallel,p_\perp) = 
2  e^{- \frac{\vert p_\perp \vert^2} {|q_{f} B|} } S_{1+1} (p_\parallel)\, ,
\end{eqnarray}
where $p_\parallel^\mu = (p^0,0,0,p^3) $ and $ p_\perp^\mu = (0, p^1, p^2,0)$. 
This of course is anticipated from the dimensional reduction in the LLL. 
The longitudinal part $S_{1+1} (p_\parallel) $ is the (1+1)-dimensional fermion propagator 
in a medium. At this moment, its explicit form is not important. 
The integration over the transverse momentum $p_{\perp}$ in Eq.~\eqref{G-Fourier} 
can be easily performed, and we then arrived at 
%\com{Factor and sign corrected.}
\begin{eqnarray}
\label{eq:Gresult}
G_\R^{0, 02} (q) 
&=& i s_f   \frac{|q_{f}B|}{8\pi }\, \(  q^{1} + i s_f  q^{2} \)
\Pi_{R}^{ 00} (q_\parallel)\, ,
\end{eqnarray}
where $ q^1$ and $q^2$ are components of the external momentum $ q^\mu$.

Remarkably, we find that the retarded Green's function $G^{0,02}_{R}$, 
which determines the medium's response to the vorticity in (3+1) dimension, 
is connected to the polarization tensor in (1+1) dimension:
%we find that the longitudinal part results 
%in the retarded current correlator of the Schwinger model 
\begin{eqnarray}
\hspace{-0.5cm}
\label{eq:pol}
i \Pi_{R}^{00} \! (q_\parallel)  \! \equiv \!  %- (ig)^2
\int \!\! \frac{d^2p_\parallel}{(2\pi)^2}  
\tr_{\scriptscriptstyle 2D}[  \gam^0  S_{1+1}(p_\parallel+q_\parallel) \gam^0  S_{1+1}(p_\parallel) ] 
 . 
\end{eqnarray}
%where $ g = 1$ in the present case. 
Furthermore, {\it since this polarization tensor $ \Pi_{R}^{00}$ is related to the chiral anomaly in (1+1) dimension, 
it is one-loop exact and is not subject to any temperature or density correction 
for the massless fermion} \cite{DJ, *Baier:1991gg, FHYY}. 
%indicating that the transport coefficient are also endowed with these properties. 
Here, 
the one-loop exact form is given by 
\begin{eqnarray}
\label{eq:pi00}
 \Pi_{R}^{00} (q_\parallel)
&=&
% -\frac{1}{\pi} \Big( \frac{q^{2}_{z}}{\o^{2}-q^{2}_{z}} \Big)\, ,
 -\frac{1}{\pi} \Big[ \frac{\(q_{3}\)^{2}}{\o^{2}-\(q_{3}\)^{2}} \Big]\, ,
%(q_\parallel^2 g_\parallel^{\mu\nu} - q_\parallel^\mu q_\parallel^\nu)
%\ \to \ \frac{i}{\pi} 
\end{eqnarray}
where $ \omega \equiv q^0$.
% and $q_z \equiv q^3 $. 
%where $ g_\parallel^{\mu\nu} = {\rm diag} (1,0,0,-1)$. 
By plugging the result of the Green's function (\ref{eq:Gresult}) and (\ref{eq:pi00}) 
into the Kubo formula \eqref{kubo}, the transport coefficient $\l$ is finally obtained as  
\begin{equation}
\label{eq:result}
\l =  \frac{ C_{A} }{2} q_f B \, . 
\end{equation}
%\com{[Minus sign removed.]}
Inserting $ \l$ into Eq.~(\ref{pertub}), we indeed verify Eq.~(\ref{jV}) which was 
also obtained from the physical argument presented in the previous section. 
We also note that one should take the $\o\to 0$ limit first in Eq.~\eqref{kubo} 
as in the perturbative computations of other vorticity-induced transport phenomena \cite{Amado:2011zx,*Landsteiner:2011cp} 
(see also Ref.~\cite{Satow:2014lva} for discussions).

The existence of $j^{0}_{V}$ also implies a corresponding term 
in the axial current $\vj_{A}=\bar \Psi (x) \gam^\mu \gam^5  \Psi (x)$.
%\com{I put the equation in line}
%\begin{eqnarray}
%j_\A^\mu (x) = \bar \Psi (x) \gam^\mu \gam^5  \Psi (x)
%\, .
%\end{eqnarray}
This is due to the relation between the vector and axial currents in the LLL, 
$  j_A^\mu = - s_f \epsilon_\parallel^{\mu\nu} j_{V \nu}$ 
with $\epsilon_\parallel^{03} = - \epsilon_\parallel^{03} = +1 $. 
From this relation $ j_A^3 = s_f j_V^0$ and the vector charge density (\ref{jV}), 
we also verify Eq.~(\ref{jA}). 
Of course, one can reach the same conclusion by starting out from Eq.~(\ref{pertub}) 
with the replacement of $ j_V^\mu$ by $j_A^\mu $.  

%The coefficient for the axial current is readily obtained as 
%\begin{eqnarray}
%\lambda_A = s_f \lambda_V = - \frac{ C_{A}}{2}   \vert q_f B \vert
%\, ,
%\end{eqnarray}
%which also verifies the result shown in Eq.~(\ref{jA}). 

We have thus far considered a single-flavor and color-neutral fermion. 
Since the flavor dependence appears only in the overall factor of $ q_f$, 
extension to multi-flavor cases is simply implemented as the sum of fermion charges. 
The color factor $ N_c$ for quarks should be included just  
as the overall factors in Eqs.~(\ref{jV}) and (\ref{jA}).

\sect{Summary and applications} 
We investigated novel anomalous transport phenomena in a chiral fluid
in the presence of both vorticity and magnetic field. 
Our main results are summarized in Eqs.~\eqref{jV} and \eqref{jA}. 
%which are consistently obtained from 
%an argument for the adiabatic level shift and a diagrammatic computation by the Kubo formula.  
%our findings with physical picture and confirmed them by field-theoretical computation.
%by the weak-coupling theory 
Our analyses suggest that the corresponding transport coefficients are, due to the relation to %the quantum anomaly, 
the chiral anomaly in (1+1) dimension, 
protected from temperature and density corrections. 
%Therefore, our results should be independent of the microscopic details of the system. 
The factorization in Eq.~(\ref{eq:Gresult}) plays a crucial role 
for establishing the relation to the chiral anomaly. 
It would be interesting to examine Eqs.~\eqref{jV} and \eqref{jA} by different approaches, 
for example, 
%by solving the Dirac equation in a rotating frame~\cite{Chen:2015hfc} 
%and by studying a strongly correlated chiral fluid~\cite{Landsteiner:2011iq}. 
by means of the analytic solution of the Dirac equation in a rotating frame~\cite{Chen:2015hfc}, 
the holographic correspondence~\cite{Landsteiner:2011iq}, 
and the Wigner function formalism~\cite{Gao:2012ix, *GW}. 

It is important to implement our findings into 
the ``anomalous magnetohydrodynamics'' \cite{Giovannini:2013oga, *Giovannini:2016whv, *Yamamoto:2015gzz}. 
Casting \eqref{jV} into a covariant form, 
we propose the following realization of the magneto-vorticity coupling 
in the framework of anomalous magnetohydrodynamics:
\begin{eqnarray}
\label{hydro}
%n\equiv 
u_{\mu} j^{\mu}_{V}
&=& n_{0}(T, \mu; B)+ \Delta n\, ,
\qquad
\Delta n=  C_{A}\omega_{\mu} B^{\mu}\, . 
%\\
%j^{\mu}_{A} = C_{A}\(\mu_{V}+ \Delta \mu_{V}\)\, B^{\mu}\, , 
%\qquad
%\Delta \mu_{V}
\end{eqnarray}
As in the conventional cases, $n_{0}$ denotes the local equilibrium density as a function of temperature $T$ 
and chemical potential $\mu$ in the absence of vorticity, and $u_{\mu}$ is the flow velocity. 
The magneto-vorticity coupling is included in $\Delta n$ %which is the covariant form deduced from Eq.~\eqref{jV} 
with $\omega^{\mu}\equiv\frac{1}{2}\epsilon^{\mu\nu\a\b}u_{\nu}\pd_{\a}u_{\b}$ 
and $B^{\mu}\equiv\tilde{F}^{\mu\nu}u_{\nu}$. 
%\changed{
As mentioned in the Introduction, 
%$\Delta n $ is regarded as 
%a higher-order correction in the conventional gradient expansion \cite{Son:2009tf}. 
%However, in the strong magnetic field limit, 
this coupling term becomes comparable in 
magnitude to the first-order terms in the presence of strong magnetic fields. 
Therefore, the modification \eqref{hydro} should be included in anomalous magnetohydrodynamics 
together with the anomalous terms already considered in Ref.~\cite{Son:2009tf}. 
%}

%In the pioneering hydrodynamic analysis with the anomaly~\cite{Son:2009tf}, 
%both vorticity and magnetic field are accounted as the first order in the gradient expansion. 
%Consequently, the coupling between the magnetic field and vorticity is dropped 
%as higher-order effects in the gradient expansion in that framework. 
%However, in the context of magnetohydrodynamics, the strength of magnetic field $\vB$ can be much larger than the gradients, 
%implying the importance of going beyond the conventional gradient expansion. 
%Therefore, the modification \eqref{hydro} should be encoded into chiral magnetohydrodynamics 
%together with the anomalous terms already considered in Ref.~\cite{Son:2009tf}. 

Finally, turning to phenomenological applications of our work, 
%It would be interesting to study the possible roles in the turbulence of chiral magnetohydrodynamics relevant to the evolution of the primordial magnetic field of the Universe.
we call attention to the relativistic heavy-ion collisions 
where both a strong magnetic field and a rotation of the quark-gluon plasma are created~\cite{Kharzeev:2015znc}.
While effects of the magnetic field and vorticity have been considered separately in the heavy-ion phenomelogy, 
%our findings clearly open a new avenue to explore 
their interplay has %not been explored 
been overlooked up to now. 
It is also interesting to investigate effects of the coupling between magnetic fields and rotations of compact stars. 

\section{acknowledgement}
We would like to thank  
Y.~Hidaka, Y.~Hirono, M.~Hongo,  D.~Kharzeev, L.~McLerran, R.~Pisarski, 
P.~Surowka, M.~Stephanov, and H.-U.~Yee for helpful discussions, 
and K.~Fukushima, X.-G.~Huang, S.~Lin, D.~Satow and N.~Yamamoto for comments on the draft.  
Y.Y. also acknowledges the stimulating environment of the ``QCD Chirality'' workshop 2016 at UCLA. 
This work is supported in part by China Postdoctoral Science Foundation under Grant No.~2016M590312 
and, at the early stage, by Japan Society for the Promotion of Science Grants-in-Aid No.~25287066 (K.H.). 
K.H. is also grateful for support from RIKEN-BNL Research Center. 
This material is based upon work supported in part by the U.S. Department of Energy, Office
of Science, Office of Nuclear Physics, under Contract No.~DE-SC0012704, and within
the framework of the Beam Energy Scan Theory (BEST) Topical Collaboration (Y.Y.).

%\begin{appendix}
%
%\section{Trace}
%\end{appendix}

\bibliography{vorticity_B}

\end{document}